\documentclass[reprint,aps,amsmath,amssymb,twoside,prd,showkeys,superscriptaddress,onecolumn]{revtex4-1}
\usepackage{verbatim}
\usepackage{comment}
\usepackage{graphicx}
\usepackage[T1]{fontenc}
\usepackage{epsfig}
\usepackage{bm}
\usepackage{amssymb}
\usepackage{float}
\usepackage{amsmath}
\usepackage{subfigure}
\usepackage{dcolumn}
\usepackage{cancel}
\usepackage[colorlinks]{hyperref}
\usepackage[usenames,dvipsnames]{color}
\hypersetup{
     breaklinks=true,
    pdfstartview={FitH},    
    colorlinks=true,       
    linkcolor=blue,          
    citecolor=red,        
    filecolor=magenta,      
    urlcolor=blue,           
    anchorcolor=green,      
    linktocpage=true
}

\def\doi{http://doi.org}

\newcommand{\HCd}{\mathcal{H}}

\def\HCdt0{\tilde{\HCd}_{0}}

\newcommand{\afffias}{Frankfurt Institute for Advanced Studies (FIAS), Ruth-Moufang-Strasse~1, 60438 Frankfurt am Main, Germany}
\newcommand{\affbgu}{Physics Department, Ben-Gurion University of the Negev, Beer-Sheva 84105, Israel}
\newcommand{\affbahamas}{Bahamas Advanced Study Institute and Conferences, 4A Ocean 
Heights, Hill View Circle, Stella Maris, Long Island, The Bahamas}

\begin{document}

\title{Lorentzian Quintessential Inflation \footnote{This essay is awarded second prize in the 2020 Essay Competition of the Gravity Research Foundation.}}
\author{David Benisty}
\email{Corresponding Author: benidav@post.bgu.ac.il}
\affiliation{\affbgu}\affiliation{\afffias}
\author{Eduardo I. Guendelman}
\email{guendel@bgu.ac.il}
\affiliation{\affbgu}\affiliation{\afffias}\affiliation{\affbahamas}
\date{23.03.2020}
\begin{abstract}
From the assumption that the slow roll parameter $\epsilon$ has a Lorentzian form as a function of the e-folds number $N$, a successful model of a quintessential inflation is obtained. The form corresponds to the vacuum energy both in the inflationary and in the dark energy epochs. The form satisfies the condition to climb from small values of $\epsilon$ to $1$ at the end of the inflationary epoch. At the late universe $\epsilon$ becomes small again and this leads to the Dark Energy epoch. The observables that the models predicts fits with the latest Planck data: $r \sim 10^{-3}, n_s \approx 0.965$. Naturally a large dimensionless factor that exponentially amplifies the inflationary scale and exponentially suppresses the dark energy scale appears, producing a sort of {\it{cosmological see saw mechanism}}. We find the corresponding scalar Quintessential Inflationary potential with two flat regions - one inflationary and one as a dark energy with slow roll behavior.
\end{abstract}
\maketitle
\section{Introduction}
The inflationary paradigm is considered as a necessary part of the standard model of 
cosmology, since it provides the solution to the fundamental puzzles of the old Big 
Bang theory, such as the horizon, the flatness, and the monopole problems
\cite{Guth:1980zm,Guth:1982ec,Starobinsky:1979ty,Kazanas:1980tx,Starobinsky:1980te,Linde:1981mu,Albrecht:1982wi,Barrow:1983rx,Blau:1986cw}. It can be achieved through various mechanisms, for instance through the introduction of a scalar inflaton field \cite{Barrow:2016qkh,Barrow:2016wiy,Olive:1989nu,Linde:1993cn,Liddle:1994dx,Germani:2010gm,Kobayashi:2010cm,Feng:2010ya,Burrage:2010cu,Kobayashi:2011nu,Ohashi:2012wf,Cai:2014uka,Kamali:2016frd,Benisty:2017lmt, Dalianis:2018frf,Dalianis:2019asr}. Almost twenty years after the observational evidence of cosmic acceleration the cause of this phenomenon, labeled as dark energy", remains an open question which challenges the foundations of theoretical physics: why there is a large disagreement between the vacuum expectation value of the energy momentum tensor which comes from quantum field theory and the observable value of dark energy density \cite{Weinberg:1988cp,Lombriser:2019jia,Merritt:2017xeh}. One way to parametrize dynamical dark energy uses a scalar field, the so-called quintessence model for canonical scalar fields \cite{Ratra:1987rm,Caldwell:1997ii,Benisty:2018qed}. In such a way that the cosmological constant gets replaced by a dark energy fluid with a nearly constant density today \cite{Zlatev:1998tr,Caldwell:1999ew,Chiba:1999ka,Bento:2002ps,Tsujikawa:2013fta}. For the slow roll approximation the scalar field behaves as an effective dark energy. The form of the potential is clearly unknown and many different potentials have been studied and confronted to observations.

These two regimes of accelerated expansion are treated independently. However, it is both tempting and economical to think that there is a unique cause responsible for a quintessential inflation
\cite{wetterich,murzakulov-etal,Guendelman:2016kwj,Guendelman:2017mkj,Guendelman:2015mva,Hossain:2014xha,Hossain:2014ova,Hossain:2014zma,Geng:2015fla,Geng:2017mic,Kaganovich:2000fc,Hossain:2014xha,Hossain:2014coa} which refers to unification of both concepts using a single scalar field. Consistency of the scenario demands that the new degree of freedom, namely the scalar field, should not interfere with the thermal history of the Universe, and thereby it should be ``invisible'' for the entire evolution and reappear only around the present epoch giving rise to late-time cosmic acceleration.
  \begin{figure}[t!]
 	\centering
\includegraphics[width=0.7\textwidth]{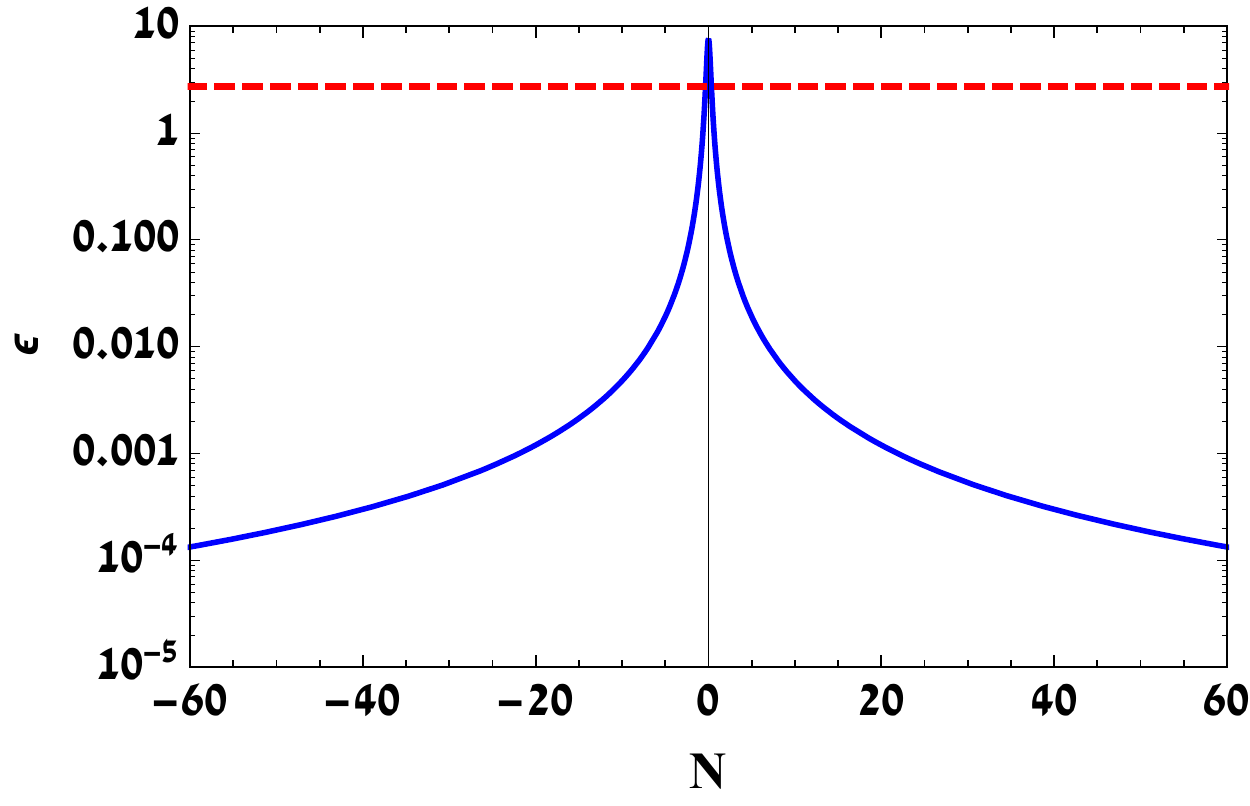}
\\
\includegraphics[width=0.7\textwidth]{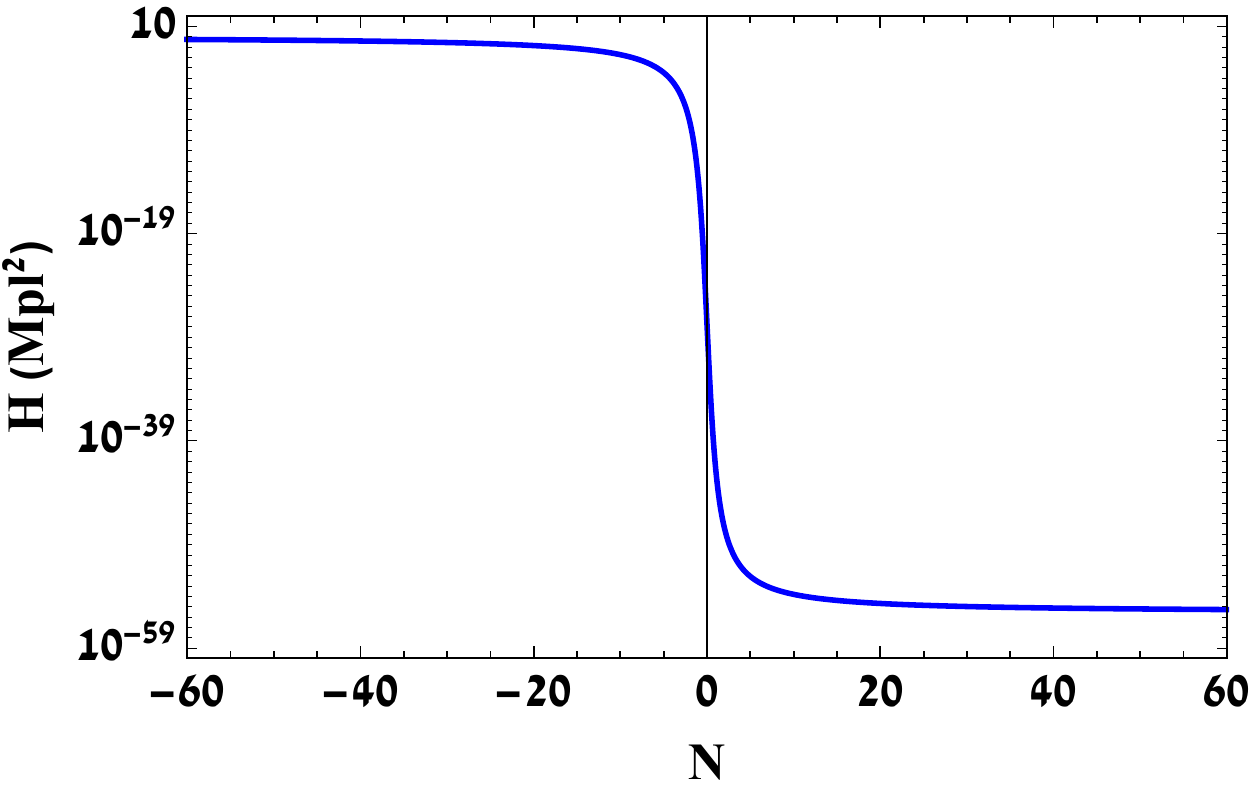}
\caption{{\it{The upper panel shows the slow roll parameter $\epsilon$ vs. the number of $e$-folds for the anzats (\ref{eq:ans}), in a logarithmic scale. The lower panel shows the corresponding Hubble function of the vacuum vs. the number of $e$-folds.}}}
\label{fig1}
\end{figure}
\section{Lorentzian Anzats}
\label{sec:LorAnz}
In order to formulate an anzats for the Hubble function that treats symmetrically both the early and late times we use the Lorentzian function for the slow roll parameter:
\begin{equation}
\epsilon(N) = \frac{\xi}{\pi}\frac{\Gamma/2 \, }{ N^2 + (\Gamma/2)^2}
\label{eq:ans}
\end{equation}
as a function of the number of $e$-folds $N = \log(a/a_i)$, where $a_i$ is the scale parameter at some time (which we may choose as the initial state of the inflationary phase). $\xi$ is the amplitude of the Lorentzian, $\Gamma$ is the width of the Lorentzian. In that way the $\epsilon$ parameter increases from the initial value to $1$ at the end of inflation,then continues to increase, peak and then decreases until it gets down to the value $1$ and this represents the beginning of a the new Dark Energy phase that will eventually dominate the late evolution of the Universe. The upper panel of Fig \ref{fig1} presents the qualitative shape of this behavior.

The strong energy condition yields another bound on the coefficients. The equation of states $w$ is in the range $ |w| \leq 1$. From the relation $\epsilon = \frac{3}{2}\left(w + 1\right)$ we obtain the bound $ 0 \leq \epsilon \leq 3$. The anzats for the vacuum energy evolution (\ref{eq:ans}) positive always, hence the lower bound is preserved. The largest value of the anzats (\ref{eq:ans}) is $2 \xi /\pi  \Gamma $. From the the upper bound of $\epsilon$ we obtain the condition:
\begin{equation}
\Gamma < 2 \xi/ 3\pi.
\label{eq:cond}
\end{equation}
In general, the calculation of the above observables demands a detailed perturbation analysis. Nevertheless, one can obtain approximate expressions by imposing the slow-roll assumptions, under which all inflationary information is encoded in the slow-roll parameters. In particular, one first introduces \cite{Martin:2013tda} 
\begin{equation}\label{slowRoll}
  \epsilon_{n+1} = \frac{d}{dN} \log 
|\epsilon_n|,
\end{equation}
where $\epsilon_0\equiv H_{i}/H$ and $n$ a positive integer. The slow roll parameters read:
\begin{equation*}
\epsilon \equiv \epsilon_1 = -\frac{H'}{H},\quad \epsilon_2 = \frac{H''}{H'}-\frac{H'}{H},
\end{equation*}
and so on. From the first slow roll parameter definition with the anzats (\ref{eq:ans}), we obtain the solution:
\begin{equation}
H = \sqrt{\frac{\Lambda_0}{3}} \exp[-\frac{\xi}{\pi}\tan ^{-1}\left(\frac{2 N}{\Gamma }\right)]. 
\end{equation}
where $\Lambda_0$ is an integration constant. The Hubble function interpolates from the inflationary values $H_{-\infty}$ to the dark energy value $H_{+\infty}$ that corresponds to:
\begin{equation}
H_{\pm } = \sqrt{\frac{\Lambda_0}{3}}\exp^{\mp \xi/2}.
\end{equation}
The magnitude of the vacuum energy at the inflationary phase reads $10^{-8} Mpl^4$, while the magnitude of the vacuum energy at the present slowly accelerated phase of the universe is $10^{-120} Mpl^4$. From the Friedmann equations the values of the energy density is $3H^2$ in the Planck scale. Therefore, the coefficients of the model are:
\begin{equation}
\xi \approx 129,\quad \Lambda_0 = 1.7 \cdot 10^{-32} Mpl^4.
\end{equation} 
We calculate the other slow roll parameters using (\ref{slowRoll}):
\begin{equation}
\epsilon_2 = -\frac{8 N}{\Gamma ^2+4 N^2}, \quad
\epsilon_3 = \frac{1}{N}-\frac{8 N}{\Gamma ^2+4 N^2}.
\end{equation}
For $\Gamma \rightarrow 0$ all of the slow roll parameters with $n \geq 3$ yields the value $-1/N$. However in the general case, all of the slow parameters have small values if the $\epsilon_2$ is small.

As usual inflation ends at a scale factor $a_f$
where $\epsilon_1(a_f)=1$ and the slow-roll approximation breaks down. Therefore the end of inflation takes place when the number of $e$-folds read:
\begin{equation}
N_{f} =  \pm \sqrt{\frac{\Gamma}{4 \pi } (2 \xi-\pi  \Gamma)}
\end{equation}
Notice that with the condition (\ref{eq:cond}) the gets a definite value. In order to have an inflationary phase the condition $2\xi > \pi \Gamma$ must be satisfied. The negative value of $N_{f}$ is the final state of the inflationary phase, while the positive value of $N_f$ is the initial value of the slow rolling Dark Energy at the late universe. Therefore, in order to calculate the inflationary observables, we must take the minus sing of $N_{f}$. we take Consequently the initial $N_i$ satisfies the condition: $N_f - N_i = \mathcal{N} \approx 50-60$,
where we impose $60\,e$-folds for the inflationary phase. Hence, the initial state of the inflationary phase reads:
\begin{equation}
N_i = - \sqrt{\frac{\Gamma}{4 \pi } (2 \xi-\pi  \Gamma)} - \mathcal{N}
\end{equation}
The inflationary observables are expressed as \cite{Martin:2013tda}:
\begin{equation}
r \approx 16\epsilon_1 , \quad n_\mathrm{s} \approx 1-2\epsilon_1-\epsilon_2, \quad \alpha_\mathrm{s} \approx -2 \epsilon_1\epsilon_2-\epsilon_2\epsilon_3  , \quad
 n_\mathrm{T} \approx -2\epsilon_1 ,   
\end{equation}
where all quantities are calculated at $N_{i}$. Therefore the tensor to scalar ratio and the primordial tilt give:
\begin{subequations}
\begin{equation}
r = \frac{32 \Gamma  \xi }{\pi  \Gamma ^2+4 \pi  N_i^2}, \quad n_s = \frac{\pi  \left(\Gamma ^2+4 N_i (N_i+2)\right)-4 \Gamma  \xi }{\pi  \left(\Gamma ^2+4 N_i^2\right)}.
\end{equation}
\end{subequations}
For $60\, e$-folds and $\Gamma = 0.1$ the observables read:
\begin{equation}
r = 0.0076, \quad n_s = 0.961754.
\end{equation}
These values in agreement with the latest $2018$ Planck data \cite{Aghanim:2018eyx,Akrami:2018odb}:
\begin{equation}
0.95 < n_s < 0.97, \quad r < 0.064
\end{equation}
Fig \ref{fig2} shows the predicted distribution of the observables \cite{getDist}. Fig \ref{fig3} shows the predicted distribution of the observables \cite{getDist}. We assume a uniform prior: $N \in [50;70]$, $\xi \in [100;200]$, $\Gamma \in [0;1]$, with $10^7$ Markov Chain Monte Carlo samples. We find the posterior yields: 
\begin{equation}
r = 0.045^{+0.065}_{-0.053}, 
\end{equation}
\begin{equation}
n_s = 0.9624^{+0.0087}_{-0.011},
\end{equation}
\begin{equation}
\alpha_s = -\left(33^{+27}_{-30}\,\right)\cdot 10^{-5}, 
\end{equation}
in a good agreement with the recent Planck values.
\section{Scalar field dynamics}
  \begin{figure*}
 	\centering
 	\includegraphics[width=0.8\textwidth]{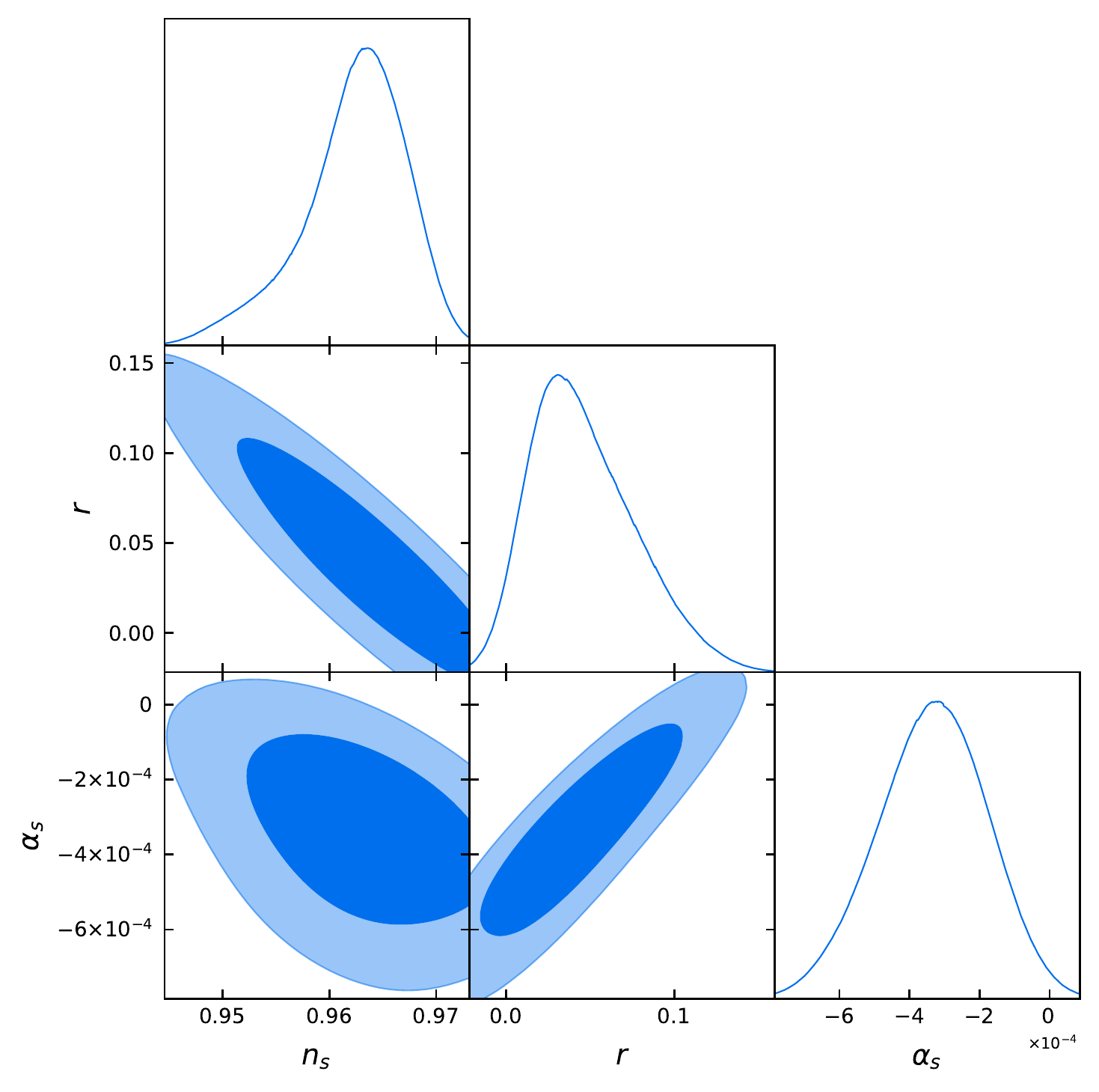}
\caption{{\it{The predicted scalar to tensor ratio vs. the primordial tilt of the model.}}}
\label{fig2}
\end{figure*}
The above anzats is of general applicability in any inflation realization, whether this is driven by a scalar field, or it arises effectively from modified gravity, or from any other mechanism. In order to provide a more transparent picture let us consider a realization of these ideas in the context of a canonical scalar field theory $\phi$ moving in a potential $V\left(\phi \right)$. The Friedmann equations are:
\begin{equation}\label{FriedInf}
 H^2  = \frac{8\pi G}{3}\left[\frac{1}{2} \dot{\phi}^2 + V(\phi)\right], \quad
 \dot{H} = -4\pi G\dot{\phi}^2,
\end{equation}
while the variation for the scalar field is
\begin{equation}
\label{KleinG}
\ddot{\phi} + 3 H \dot{\phi} + V'(\phi) = 0.
\end{equation}
Let us apply the anzats in order to reconstruct a physical scalar-field potential that can generate the desirable inflationary observables. From the Friedmann equation (\ref{FriedInf}) that holds in every scalar-field inflation, we extract the 
following solutions:
\begin{equation}\label{conTran1}
\phi = \int_{0}^{N} \sqrt{-2 \frac{H'}{H}} \, dN, \quad
V(\phi) = H H'+3 H^2.
\end{equation}
with $8\pi G = 1$. From the integration of the Hubble parameter we get:
\begin{equation}
N = \frac{\Gamma}{2}   \sinh \left(\sqrt{\frac{\pi}{\xi \Gamma}}\phi\right), \quad 
V(N) = \Lambda_0 e^{-\frac{2 \xi}{\pi }\tan^{-1}\left(\frac{2 N}{\Gamma }\right)} \left(1-\frac{2 \Gamma  \xi }{3 \pi  \Gamma ^2+12 \pi  N^2} \right).     
\label{VphiN}
\end{equation}
Expression (\ref{VphiN}) cannot be inversed, in order to find $N(\phi)$ and then through 
insertion into (\ref{VphiN}) to extract $V(\phi)$
analytically:  
\begin{equation}
\begin{split}
V(\phi) = \Lambda_0 e^{-\frac{2 \xi}{\pi} \tan ^{-1}\left(\sinh x \right)} \left(1-\frac{2 \xi }{3 \pi  \Gamma } \text{sech}^2 x \right). 
\end{split}
\label{FinPot}
\end{equation}
with $x \equiv \sqrt{\pi/\Gamma \xi} \phi$. Fig \ref{fig2} shows the scalar potential $V(\phi)$. The universe in this picture begins with $\phi \rightarrow \infty$ with a slow roll behavior and goes to the left-hand side. After approaching the minimum the universe evolves with another slow roll behavior that corresponds to the dark energy epoch when $\phi \rightarrow  - \infty$. The asymptotic values of the potential are:
\begin{equation}
V_{+\infty} = \Lambda_0 e^{\xi }, \quad V_{-\infty} = \Lambda_0 e^{-\xi } .
\end{equation}
Notice that this represents a see saw cosmological effect, that is if $\Lambda_0$ represents an intermediate scale, we see that in order to make the inflationary scale big forces the present vacuum energy to be small.  $\Lambda_0$ represents the geometric average of the inflationary vacuum energy and the present Dark Energy vacuum energies.
  \begin{figure}[ht]
 	\centering
\includegraphics[width=0.7\textwidth]{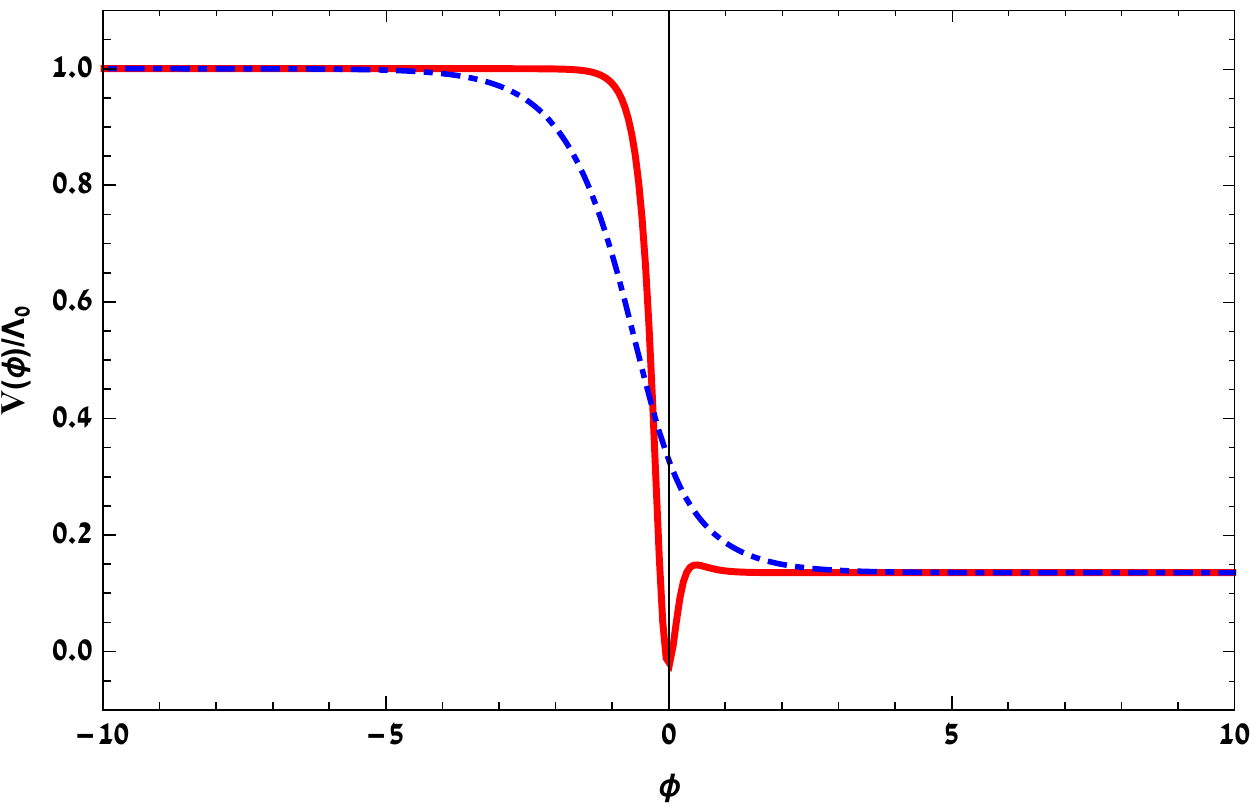}
\caption{{\it{The corresponding scalar field potential for the Lorenzian anzats, with different values of $\Gamma$: $0.1$ red smooth line, $1$ blue dashed line.}}}
\label{fig3}
\end{figure}
\section{Discussion}
This essay introduces a model where we start with an ansatz for the slow roll parameter $\epsilon$ for the whole history of the Universe . We choose a Lorentzian form for $\epsilon$, which peaks at some point and goes to zero for the early and late Universe, so these two epoch have an accelerated phase. The magnitude of the vacuum energies at the early and late Universe obeys a see saw mechanism, since the asymptotic values of the potential are $\Lambda_0 e^{\pm \xi }$ represents a see saw cosmological effect, where the requirement that one scale (the inflationary scale) be large pushes the Dark Energy scale to be very low. The magnitude of the vacuum energies at the early and late Universe obey a see saw mechanism, since the asymptotic values of the potential are $\Lambda_0 e^{\pm \xi }$ representing a see saw cosmological effect, where the requirement that one scale (the inflationary scale) be large pushes the Dark Energy scale to be very low. See saw cosmological effects in modified measure theories with spontaneously broken scale invariance have been studied in \cite{Guendelman:1999rj,Guendelman:1999qt,Guendelman:2014bva}.
For the situation presented in this paper to work, we must choose $\Lambda_0$ as an intermediate scale, and indeed then we see that in order to make the inflationary scale big, this  forces the present vacuum energy to be small.  $\Lambda_0$ represents the geometric average of the inflationary vacuum energy and the present Dark Energy vacuum energies.

The model formulates the vacuum energies both in the inflationary epoch and in the dark energy epoch. However to compare the basis of the model with the whole history of universe, we have take into account particle creation models with temperature, as well as radiation production.

\section*{Acknowledgments}
This article is supported by COST Action CA15117 "Cosmology and Astrophysics Network for Theoretical Advances and Training Action" (CANTATA) of the COST (European Cooperation in Science and Technology). This project is supported by COST Actions CA16104 and CA18108.  D.B. and E.I.G thanks FQXi and the Ben-Gurion University of the Negev for great support. D.B. thanks to Frankfurt Institute for Advanced Studies for generous support. D.B. thanks to Bulgarian National Science Fund for support via research grant KP-06-N 8/11.

\bibliography{ref}
\end{document}